\documentclass[twocolumn]{aa} 
\usepackage[pdftex]{graphicx}
\usepackage{txfonts}
\begin{document}
\title{Kinematics and H$_2$ 
morphology of the multipolar Post-AGB star
IRAS 16594$-$4656\thanks{Based on observations collected at the
Gemini-South with the Phoenix instrument under GS-2003A-Q-41, and VLT under 65.L-0615(A) }}  

\subtitle{}

\author{G. C. Van de Steene\inst{1}
\and T.\ Ueta\inst{1,2}
\and P.\ A.\ M.\ van Hoof \inst{1}
\and M. Reyniers\inst{3}
\and A.\ G.\ Ginsburg\inst{4}}

\offprints{G. C. Van de Steene}

\institute{
Royal Observatory of Belgium, Ringlaan 3, 1180 Brussels, Belgium \\
\email{gsteene@oma.be} \\ \email{pvh@oma.be}
\and Department of Physics and Astronomy, University of Denver, 2112 E. 
Wesley, Denver, CO 80208, U.S.A. \\     
\email{tueta@du.edu}
\and Instituut voor Sterrenkunde, K.U.Leuven, Celestijnenlaan 200D,  B-3001 Leuven \\
\email{maarten@ster.kuleuven.ac.be}
\and Department of Astrophysical and Planetary Sciences, University of  Colorado, Boulder,
 CO 80309, USA
}

\authorrunning{Van de Steene et al.}

\titlerunning{H$_2$ emission in IRAS 16594$-$4656}

\date{Received / Accepted}

\abstract 
{
The spectrum of IRAS 16594$-$4656 shows shock-excited H$_2$ 
emission and collisionally excited emission lines such as
[O\,{\sc i}], [C\,{\sc i}], and [Fe\,{\sc ii}].
} 
{The goal is to determine the location of the H$_2$ and
[Fe\,{\sc ii}] shock emission, to determine the shock velocities,
and to constrain the physical properties in the shock.  }  
{High resolution spectra of the H$_2$~1-0~S(1), 
H$_2$~2-1~S(1), [Fe\,{\sc ii}], and Pa$\beta$ emission 
lines were obtained with the near infrared
spectrograph Phoenix on Gemini South. }  
{The position-velocity diagrams of H$_2$ 1-0 S(1), H$_2$ 2-1 S(1),
and [Fe\,{\sc ii}] are presented.  The H$_2$ and [Fe\,{\sc ii}]
emission is spatially extended. 
The collisionally excited [O\,{\sc i}] and [C\,{\sc i}]
optical emission lines have a similar double-peaked profile compared to the extracted H$_2$
profile and appear to be produced in the same shock.  They all
indicate an expansion velocity of $\sim$8~km~s$^{-1}$ and the presence
of a neutral, very high-density region with $n_{\rm e}$ about 
$3\,\times\,10^6$ to $5\,\times\,10^7$\,cm$^{-3}$.  
However, the [Fe\,{\sc ii}] emission is single-peaked. It has a Gaussian FWHM of
30\,km\,s$^{-1}$ and a total width of 62\,km\,s$^{-1}$ at 1\% of the peak.
The Pa$\beta$ profile is even wider with a Gaussian FWHM of 48 km\,s$^{-1}$ 
and a total width of 75\,km\,s$^{-1}$ at 1\% of the peak.  }
{The H$_2$ emission is excited in a slow 5 to 20 km\,s$^{-1}$
shock into dense material at the edge of the lobes, caused by the interaction of 
the AGB ejecta and the post-AGB wind.  
The 3D representation of the H$_2$ data shows a hollow
structure with less H$_2$ emission in the equatorial region. The
[Fe\,{\sc ii}] emission is not present in the lobes, but originates
close to the central star in fast shocks in the post-AGB wind or in a
disk.  The Pa$\beta$ emission also appears to originate close to the
star. 

}


\maketitle

\section{Introduction}
\label{introduction}

Post-Asymptotic Giant Branch (AGB) stars represent an important transition phase in the
evolution of low and intermediate-mass stars, between the AGB and the 
planetary nebula (PN) phases. During this
period, the detached circumstellar envelope of gas and dust is
expanding away from the star.  Meanwhile the star itself is increasing
in temperature at nearly constant luminosity.  This phase lasts a few
hundred to a few thousand years depending upon the star's core mass.  When the
temperature is high enough and the star photoionizes the nebula, it
has entered the PN phase (e.g. Kwok \cite{Kwok93}).
In spite of extensive study, the evolution from the AGB toward the PN
stage is still poorly understood.  The drastic changes observed in
circumstellar structure and kinematics are particularly puzzling.
During the late AGB or early post-AGB evolutionary stages, the geometry of
the circumstellar material changes from more or less spherically
symmetric to axially symmetric, with the result that most PNe exhibit
axisymmetric structures, ranging from elliptical to bipolar 
(e.g., Balick \& Frank \cite{Balick02}).  

Bipolar PNe frequently possess molecular envelopes that are readily
detectable in the near-infrared ro-vibrational lines of H$_2$
(Kastner et al.\ \cite{Kastner96}; Kelly \& Hrivnak \cite{Kelly05}).
The available data suggest that the onset of near-infrared H$_2$
emission in PNe can be traced back to the proto-planetary nebula (PPN)
phase but not back to the AGB phase of evolution (Weintraub et al.\
\cite{Weintraub98}; Davis et al.\ \cite{Davis05}).  
These observations suggest that further studies
of H$_2$ emission from PPNe may offer insight into the transition from
AGB star to PN and from spherical to axisymmetric mass loss.  The
study of transition objects showing H$_2$ emission at an early stage
is crucial for understanding the hydrodynamic processes shaping the
nebulae. The H$_2$ lines can reveal details about the physical
conditions in the shocks associated with these hydrodynamic processes
and thereby help constrain models of the interaction of the central
star with the AGB remnant. 

IRAS 16594$-$4656 is classified as a post-AGB star for several
reasons.  In the IRAS color-color diagram it has the colors of a PN
(Van de Steene \& Pottasch \cite{VdSteene93}), but it has not been
detected in the radio continuum at 3 or 6 cm. 
It has a large infrared excess due to dust with a color temperature of
173~K. It displays a double-peaked spectral energy distribution, with
the peak in the mid-infrared much brighter than the peak in the
near-infrared (Van de Steene et al.\ \cite{VdSteene00a}).  It possesses
a CO envelope with an expansion velocity of 14 km\,s$^{-1}$ (Woods et
al.\ \cite{Woods05}).  The
chemistry of IRAS 16594$-$4656 appears to be carbon-rich. This is
based on the detection of unidentified IR emission features at 3.3,
6.2, 7.7, 8.6, 11.3, 12.6, and 13.4~$\mu$m, as well as the 21~$\mu$m
feature (Garc\'{\i}a-Lario et al. \cite{Garcia-Lario99}), all commonly
associated with a carbon-rich chemistry. The optical spectrum of
IRAS 16594$-$4656 shows a spectral type B7 with significant
reddening. The H$\alpha$ emission has
a P-Cygni type profile indicative of a stellar wind (Van de Steene et al.\
\cite{VdSteene00b}). 
In Van de Steene \& van Hoof (\cite{VdSteene03}, hereafter Paper\,I) we derived 
the total extinction value (i.e., interstellar and circumstellar extinction 
combined) using the extinction law of Cardelli et al. (\cite{Cardelli89}).
We found A$_{V}=7.5\,\pm\,0.4$\,mag with R$_V\,=\,4$.
With this extinction value and the flux calibration from the Kurucz model,
we determined a distance of ($2.2\,\pm\,0.4$)\,L$_4$$^{1/2}$\, kpc,
which was in good agreement with the distance of 2.5\,L$_4$$^{1/2}$\,kpc
derived by Su et al. (\cite{Su01}) (with L$_4$ the luminosity in units of $10^4$\,L$_{\sun}$).

In the optical the nebulosity is dominated by scattered light.  HST
WFPC2 images showed that the object is a multipolar reflection nebula
with 3 extensions on each side (e.g., Hrivnak et al.\
\cite{Hrivnak99}). It has 
been interpreted as a bipolar outflow episodically channeled from a
rotating/precessing torus. The outer nebula also shows concentric arcs
(Hrivnak et al.\ \cite{Hrivnak01}).  High resolution mid-infrared images in the N and
Q-band show a bright equatorial torus viewed almost edge-on and a pair
of bipolar lobes, which show a close correspondence with the H$_2$ HST
map. The shape of the bipolar lobes shows that the fast outflow is
still confined by the remnant AGB shell (Volk et al.\ \cite{Volk06}).
The elongation of the inner nebula is along the symmetry axis of this
torus at $\sim 80^{\circ}$~PA, and it seems likely that the material
ejection is channeled into the bipolar lobes by the torus and/or some
other collimation mechanism (Garc\'{\i}a-Hern\'andez et
al.\ \cite{Hernandez04}).  Ueta et al. (\cite{Ueta05}, \cite{Ueta07})
found, based on polarized-flux images, that the inner structure of the
nebula is clearly elongated in the east- west direction (5\farcs0 $\times$
2\farcs0 at PA~81\degr). Their polarized flux maps uncover the bipolar cusp
structure of the shell, corresponding to the main elongation in
intensity, and a hollow shell structure which delineates the wall of
the elongated bipolar cavities.

In Paper\,I we examined the near-infrared spectrum of IRAS
16594$-$4656.  It shows strong H$_2$ emission lines and some typical
metastable shock-excited lines such as [Fe\,{\sc ii}] 1.257~\&
1.644~$\mu$m. We argued that the molecular hydrogen emission is mainly
collisionally excited in a C-type shock. However, H$_2$ and [Fe\,{\sc
ii}] emission don't usually coexist.  The goal of this paper is to
clarify the location of the H$_2$ and [Fe\,{\sc ii}] shock emission in
this object and to determine the shock velocities.

High resolution near-infrared spectra of IRAS 16594$-$4656 have been
obtained with the near infrared high resolution spectrograph Phoenix
on Gemini South.  The observations and data reduction are discussed in
Section \ref{observations}. The results of the analysis are presented
in Section \ref{results}, and the morphology, kinematics,
and shock properties are discussed in Section
\ref{discussion}, and finally we present some conclusions in Section
\ref{conclusions}.

\section{Observations and data reduction}
\label{observations}

\subsection{High resolution infrared spectra}

The spectra were obtained in service mode with the near infrared high
resolution spectrograph Phoenix (Hinkle et al. \cite{Hinkle02}) on
Gemini South at 3 position angles (PA), through regions showing bright
H$_2$ emission : at 7$\degr$, 65$\degr$, and 125$\degr$. The position
angle is measured in degrees east of north. The slit positions are shown on top of the HST H$_2$ 1-0 S(1)
image in Fig. \ref{slitpos}.  The slit length is 14\arcsec\ and slit
width is 3 pixels or 0\farcs25, giving a resolution of $\sim$60,000.
On 5 May 2003 the integration times were 4~$\times$~300~s for H$_2$ 1-0 S(1)\,2.12125\, $\mu$m 
through filter K4748/4712.82 and 4~$\times$~200~s for
[Fe\,{\sc ii}]\,1.64355\,$\mu$m through filter H6073/6082.93 at each
position angle.  H$_2$ 2-1 S(1)\,2.2471\,$\mu$m was observed for 
4~$\times$~300~s at each position angle through filter K4484/4448.98 on 1 August
2003. Next, in the same night, Pa$\beta$ was observed, however the
weather conditions were quickly deteriorating and we obtained only one
useful image of Pa$\beta$ before the dome had to be closed.

In the archive there was another set of H$_2$ 1-0 S(1) data available
for this object taken with Phoenix under GS-2003A-Q-27 (PI: B.\ J.\
Hrivnak; Hrivnak et al.\ \cite{Hrivnak06}) at 33$\degr$, 52$\degr$,
72$\degr$, 345$\degr$ PA through the star, and at 345$\degr$ PA offset
$0\farcs9$ west of the star, all with a slit width of $0\farcs35$
giving a resolution of $\sim 50,000$.  We include this data set in our
analysis for the sake of completeness and in order to reconstruct the
3-dimensional structure of the H$_{2}$ shell.

Basic data reduction (such as flatfielding and sky subtraction) was
done in {\sc iraf} using the scripts provided by the Phoenix team.
Next the images were rectified by fitting the position of the
continuum peak in the spatial dimension along all pixels in the
spectral dimension.  Then the images were combined to generate cross-
correlated, co-added images.  Continuum emission was subtracted from
the data by fitting the emission of the
source in the spectral dimension before co-adding. The wavelength
calibration was done using telluric lines in the standard star
spectrum using the HITRAN 2000 database (Rothman et al. \cite{Hitran00}).  The
resulting position-velocity diagrams are presented in 
Figs.\,\ref{mapH21-0}\,-\,\ref{mapfe2}.

\subsection{H$_2$ 1-0 S(1) image}

\begin{figure}
\mbox{\includegraphics*[width=\columnwidth]{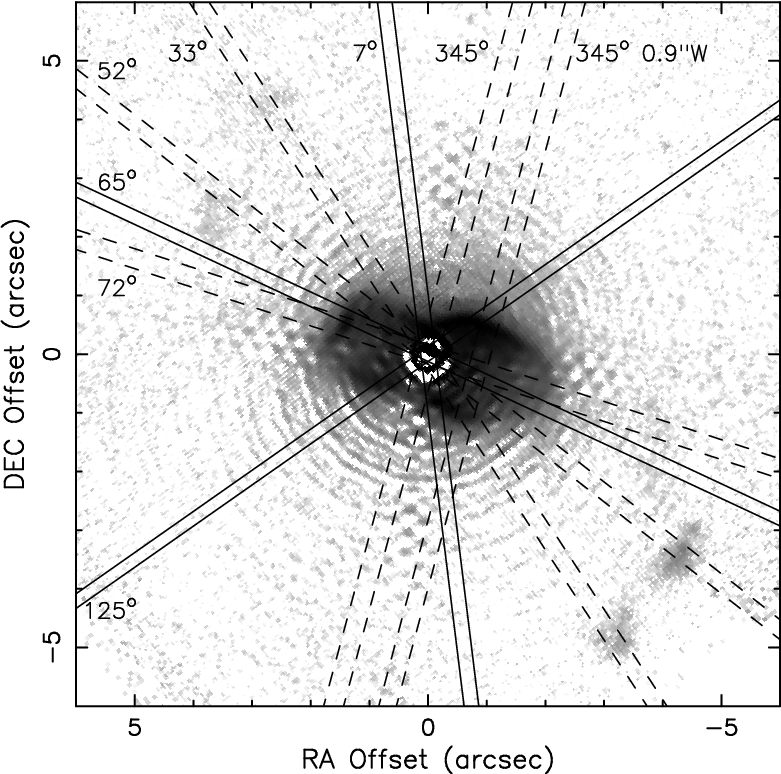}}
\caption{\label{slitpos}HST/NICMOS H$_2$ 1-0 S(1) image (north is up and east to the left) 
showing the slit positions. The slit positions displayed are at 7$\degr$, 65$\degr$, and 125$\degr$ PA with
 $0\farcs25$ width (this work; solid line) and at 33$\degr$, 52$\degr$,
 72$\degr$, 345$\degr$ PA, and 345$\degr$ PA offset $0\farcs9$ west of the star
 with $0\farcs34$ width (Hrivnak et al.\ \cite{Hrivnak06}; dashed line).
}
\end{figure}

We retreived the high-resolution H$_2$ image from the HST archive.
IRAS 16594$-$4656 was observed with {\it NICMOS} (Malhotra et al.\
\cite{nicmosihb}) on-board HST in Cycle 11 (General Observer program
9366, PI: B.\ Hrivnak) on 12 January 2003.  The observations were
made with NICMOS2, which provides a $19\farcs2 \times 19\farcs2$
field of view with $0\farcs075$ pixel$^{-1}$ scale, in conjunction with the
F212N (H$_{2}$) and F215N (H$_{2}$ and Br$\gamma$ continuum) filters.
We reduced the data using the the standard set of NICMOS calibration
programs provided in {\sc iraf/stdas} version 3.1 \footnote{STSDAS is
a product of the Space Telescope Science Institute, which is operated
by AURA for NASA}.  A detailed account of the reduction procedure can
be found in Ueta et al. (\cite{Ueta05}).  The reduced image is shown
in Fig.\ \ref{slitpos} with the slit positions of the available
Phoenix data.

\subsection{UVES spectrum}

A high resolution echelle spectrum was obtained with {\it UVES} on 
the VLT at ESO Paranal in May and June 2000. A detailed account of the 
observations, reduction procedure, and spectrum can be found in Reyniers
(\cite{Reyniers02}).

\section{Results}
\label{results}

\subsection{H$_2$ 1-0 S(1)}

\begin{figure*}
\mbox{\includegraphics*[width=18cm]{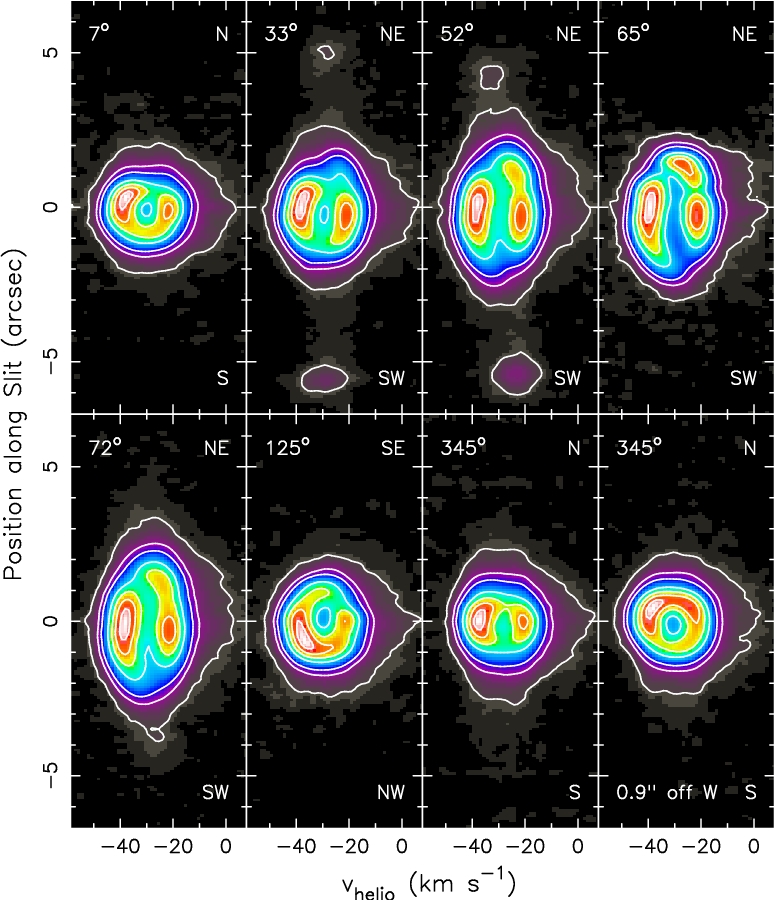}}
\\
\caption{\label{mapH21-0}%
The position-velocity diagrams of the spectra at H$_2$ 1-0 S(1).
The slit position angles and corresponding spatial directions are
indicated at the top and bottom corners.  Contours are, from the top to
 bottom, $90\%$, $70\%$, $50\%$, $30\%$, $10\%$, $5\%$, and $1\%$ of the peak.  }
\end{figure*}

The H$_2$ 1-0 S(1) position-velocity diagrams in Fig.\ \ref{mapH21-0}
can be understood in the context of the detailed morphology of H$_2$
seen in the H$_2$ NICMOS image.  The slit positions are shown on the
HST H$_2$ image (Fig.\ \ref{slitpos}).  The slit position at
7\degr\ primarily goes through the center, which is mostly
washed out in the NICMOS H$_2$ image due to the bright central star.
The slit position at 65\degr\ is almost along the major axis and going
through the bright spot at the tip of the eastern lobe.  The slit
position at 125\degr\ goes through the center at an intermediate
angle and captures the bright northwestern edge of the lobe.

Reyniers (\cite{Reyniers02}) observed a large spread in radial
velocity of the optical lines (from $-10$ to $-40$\,km\,s$^{-1}$) with a
small dependence upon excitation potential. The high excitation
emission lines seem to have mean radial velocities around
$-30$\,km\,s$^{-1}$, while the absorption lines have velocities more
around $-23$\,km\,s$^{-1}$ in the mean.  The extracted H$_2$ profiles
are double-peaked. The minimum flux in the H$_2$ profiles is on
average at $-29.2$\,km\,s$^{-1}$. Hence the H$_2$ velocity seems to
agree on average with the velocity of the emission lines in the
optical spectrum. We adopt $-29.2$\,km\,s$^{-1}$ as reference velocity.

The slit position at 7$\degr$ PA (Fig.\ \ref{mapH21-0}, top left)
shows a toroidal structure of $2\farcs4$ in diameter in the
north-south direction at $10\%$ of the peak. The diameter of the shell
extends to about $4\arcsec$ at $1\%$ ($\sim 5 \sigma$) of the peak
emission. There is little emisson around the central star, reinforcing
the toroidal appearance.  It may correspond to the toroidal equatorial 
density enhancement (EDE) as
seen in the mid-infrared images (Volk et al. \cite{Volk06}).  The bulk
of the near/approaching side of the shell moves at $-38.2$\,km\,s$^{-1}$ and the
far/receding side at $-21.2$\,km\,s$^{-1}$. Hence, with respect to the
reference velocity of $-29.2$\,km\,s$^{-1}$, the bulk of H$_2$
emission appears to expand at about 8\,km\,s$^{-1}$.  This seems low
compared to the CO (J=2-1) outflow velocity of 14\,km\,s$^{-1}$ (Woods
et al.\ \cite{Woods05}).  However, the total velocity width, including
the wings, is as large as $50.8\,(\pm\,2.2)$\,km\,s$^{-1}$ at the
$1\%$ level.  The velocity spread in the wing at the receding side is
much larger than at the approaching side.

The slit position at 65$\degr$ PA (Fig.\ \ref{mapH21-0}, top right)
samples the lobes close to the major axis.  In this spectral image the
object is most elongated: about $5.5\arcsec$ at the $1\%$ ($2 \sigma$)
level of the peak.  The bulk of the emission is detected within
$2\arcsec$ of the central star, most of it towards the northeast,
which corresponds to the main emission region in the HST H$_2$ image
(Fig.\ \ref{slitpos}).  The lack of emission at the reference velocity
close to the star points to a hollow nature of the
shell.  At this slit position, the main lobe is separated into two
emission regions that are moving away from the central region.  The
approaching peak is located almost at the center at $-39.5$\,km\,s$^{-1}$, 
while the receding shell is slightly off ($0\farcs3$) to the
southwest and moves at $-20.6$\,km\,s$^{-1}$.  The third emission peak
represents the northeastern ``tip'' at $1\farcs4$ from the centre and
moves at $-26.7$\,km\,s$^{-1}$, which is close to the reference velocity.
The approaching peak is strongest and appears to be elongated towards
the southwest. The presence of the blob at the northeastern tip seems
to make the receding side appear to be elongated towards the
northeast, giving a point-symmetric appearance to the whole shell
structure.  

At 125$\degr$ PA (Fig.\ \ref{mapH21-0}, bottom second left) the slit
still passes through the EDE, but also through the region of bright
H$_2$ emission in the western lobe.  The shell structure appears to be
hollow, which is consistent with what is seen in the data
at other slit positions.  The position-velocity diagram suggests that
most of the bright emission at the northwestern side of the lobe comes
from the approaching side of the lobe (at $0\farcs7$ to the northwest with
a velocity of $-36.8$\,km\,s$^{-1}$)

Data at other slit positions have been discussed by 
Hrivnak et al.\,(\cite{Hrivnak06}).  The overall trend of the emission structure is
similar for the two data sets at similar position angles, except
that the Hrivnak data set covers emission from the clumps located at
$\sim 5\arcsec$ from the centre. The clumps are almost at the reference
velocity. However, especially at the 52\degr\ PA, there is a slight
velocity difference between these clumps: the clump at the northeast is a bit
blueshifted and the one at the southwest is a bit redshifted compared to
the reference velocity. The slit positions at 33\degr\ and 125\degr\ PA, and 
as well as at 72\degr\ and 345\degr\ PA are almost perpendicular to each other. 
The data confirm the hollow structure which is round and elongated, but the H$_2$ intensity
is not uniformly distributed along the walls.

In general the position-velocity diagrams suggest that (1) the shell
is hollow, (2) its axis is nearly aligned with the plane of the sky, (3)
the approaching side of the shell shows stronger emission and is
elongated towards the southwest, while the position of the emission peak is slightly
northeast of the centre, and (4) the receding side of the shell is weaker
in emission strength and is elongated towards the northeast, while the
position of the emission peak is slightly southwest of the centre.  Thus the emission
structure appears to be almost point-symmetric in both the spatial and
velocity dimensions.

\subsection{H$_2$ 2-1 S(1) emission}

\begin{figure*}
\mbox{\includegraphics*[width=18cm]{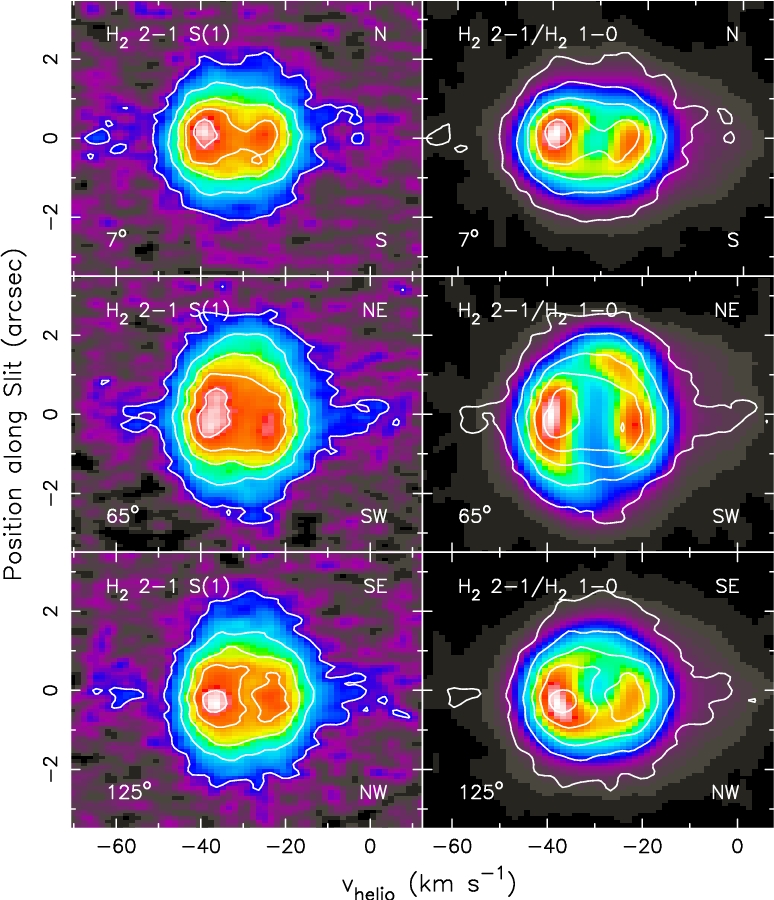}}
\\
\caption{\label{mapH22-1}%
The position-velocity diagrams of the spectra at H$_2$ 2-1 S(1) (left)
 and H$_{2}$ 2-1 S(1) contours overlaid on the H$_{2}$ 1-0 S(1) images (right).
The slit position angles and corresponding spatial directions are
indicated at the top and bottom corners.  Contours are, from the top to
 bottom, $90\%$, $70\%$, $50\%$, $30\%$, and $10\%$ of the
 peak.  When conturs are overlaid, the H$_{2}$ 1-0 S(1) images are smoothed to
 have matching spatial and spectral resolution.}
\end{figure*}

Fig.\ \ref{mapH22-1} shows the H$_2$ 2-1 S(1) position-velocity diagrams
(left column) and H$_2$ 2-1 S(1) contours overlaid on the H$_2$ 1-0 S(1)
position-velocity diagrams (right column at 7\degr, 65\degr, and 125\degr\ PA, from top to
bottom, respectively). The H$_{2}$ 2-1 S(1) and 1-0 S(1) emission
seem to be extended in a similar manner, except that (1) H$_2$ 2-1
S(1) seems to originate a bit closer to the central star than H$_2$ 1-0 S(1) and (2)
H$_{2}$ 2-1 S(1) does not show an elongation towards the northeast (in the receding
lobe) and to the soutwest (in the approaching lobe) at 65\degr\ PA.  Also, H$_{2}$ 2-1
S(1) does not seem to show the large velocity dispersion towards the red as 
in the H$_{2}$ 1-0 S(1) images. The lowest contour (at $1\%$ of the peak) displays
elongations into higher/lower velocities near the centre, but this 
is probably residual continuum emission that has not been
completely removed. 

\subsection{ [Fe\,{\sc ii}] 1.645$\mu$m emission}

\begin{figure*}
\mbox{\includegraphics*[width=18cm]{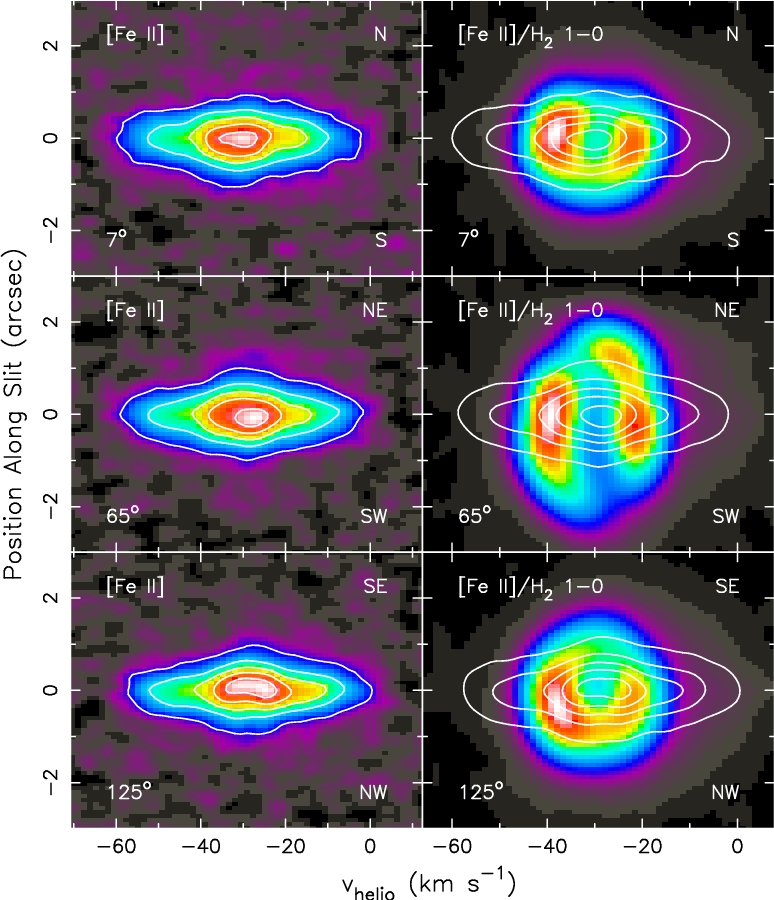}}
\\
\caption{\label{mapfe2}%
[Fe\,{\sc ii}] color images with [Fe\,{\sc ii}] contours
 overplotted (left) and [Fe\,{\sc ii}] contours overplotted 
on the H$_2$ 1-0 S(1) images (right) 
at $7\degr$, $65\degr$, and $125\degr$ PA from top to
 bottom, respectively.}
\end{figure*}

Fig.\ \ref{mapfe2} shows the [Fe\,{\sc ii}] position-velocity diagrams
(left column) and the [Fe\,{\sc ii}] contours overlaid on the H$_2$
1-0 S(1) position-velocity diagrams (right column) at 7\degr, 65\degr, and 125\degr\ PA, from
top to bottom, respectively.  From the contour-overlaid images it is
clear that the H$_2$ and [Fe\,{\sc ii}] emission originates from
different regions.  

The seeing of the night was between 0\farcs7 and 0\farcs9 in the
R-band.  Using the stellar continuum we measure a seeing of 0\farcs8,
0\farcs7, and 0\farcs6 on average at 7\degr, 65\degr, and 125\degr\
respectively.  The Gaussian FWHM of the [Fe\,{\sc ii}] profile is
0\farcs98, 1\farcs05, and 0\farcs94 at 7\degr, 65\degr, and 125\degr\
respectively.  Hence we conclude that the [Fe\,{\sc ii}] emission
region is extended. At an assumed luminosity of $10^4$\,L$_{\sun}$ and
distance of 2.2\,kpc (Paper\,I) the deconvolved Gaussian FWHM diameter of the
emitting region would be 0.004\,pc\,$\pm$\,0.001 pc (or $80\,\pm20$\,AU).
Assuming a stellar temperature of 14,000\,K (Reyniers \cite{Reyniers02}),
this is about 1800 stellar radii.

\begin{figure}
\mbox{\includegraphics*[width=\columnwidth]{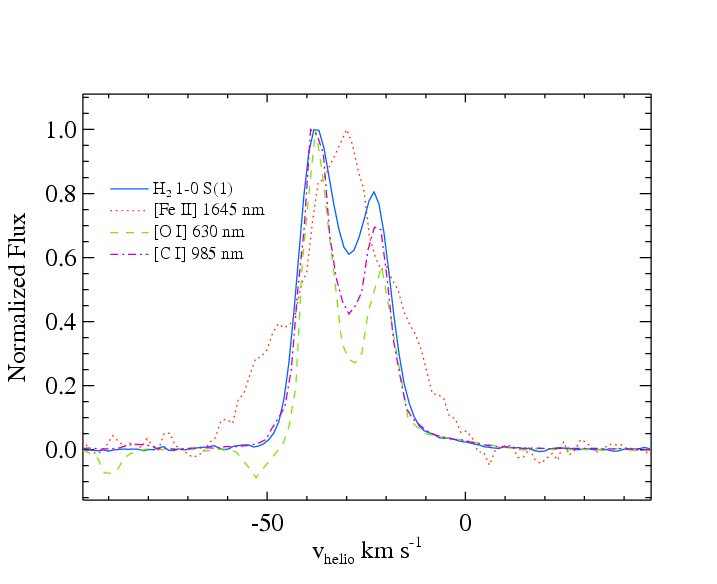}}
\caption{ Velocity profiles of the extracted spectra of 
H$_2$ 1-0 S(1) at 7\degr\ PA, [Fe\,{\sc ii}] at 7\degr\ PA, [O\,{\sc i}], and [C\,{\sc i}]. 
Each line is continuum subtracted and normalized to the peak intensity. }
 \label{velocityfig1}
\end{figure}

The extracted profile at 7\degr\ PA is shown in Fig.\,\ref{velocityfig1}.
The [Fe\,{\sc ii}] emission peaks at the reference
velocity of $-29.2$\,km\,s$^{-1}$ and the Gaussian FWHM of the profiles
are 31.8\,km\,s$^{-1}$, 33.1\,km\,s$^{-1}$, and 33.1\,km\,s$^{-1}$ at
7\degr, 65\degr, and 125\degr\ PA, respectively.  The total width of the
profile at 1\% of the peak is 62\,km\,s$^{-1}$.  The [Fe\,{\sc ii}] emission appears to
come from the inner region, close to the star at all position angles
and shows a larger velocity dispersion than the H$_2$ emission.

\subsection{Pa$\beta$ and other Paschen lines}

One useful image of Pa$\beta$ was obtained at $7\degr$ PA before the
dome had to be closed due to bad weather. The Pa$\beta$ emission is
centered on the star. The seeing was at least 1\farcs5, much worse
than for the [Fe\,{\sc ii}] observations, and the spatial profile
remained unresolved.

\begin{figure}
\mbox{\includegraphics*[width=\columnwidth]{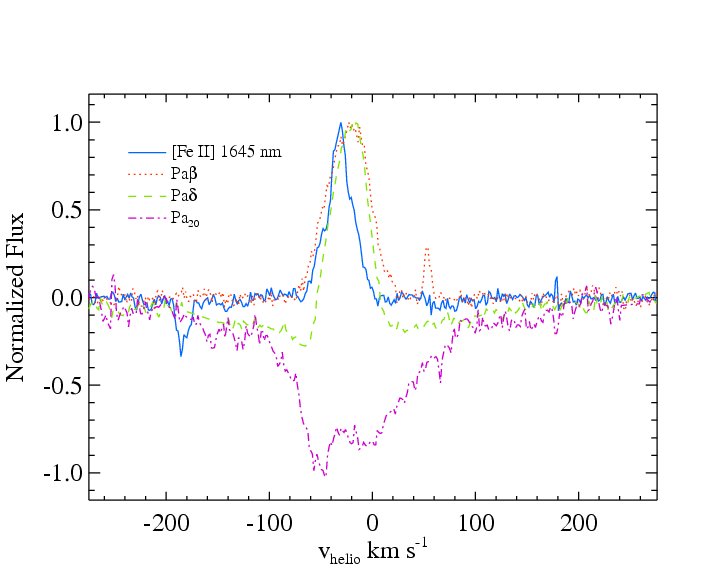}}
\caption{ Velocity profiles of the extracted spectra  of
3 Paschen lines and [Fe\,{\sc ii}]. Each line is continuum subtracted and 
normalized to the peak intensity. }
\label{velocityfig2}
\end{figure}

In Fig. \ref{velocityfig2} we compare Pa$\beta$ with the
[Fe\,{\sc ii}] profile at 7$\degr$\ PA. The peak of the Pa$\beta$ profile is
at $-21$\,km\,s$^{-1}$, while the peak of [Fe\,{\sc ii}] is at $-29.8$\,km\,s$^{-1}$.  
The velocity of the peak of the Pa$\beta$
profile is closer to the average velocity of the absorption lines in
the optical spectrum, while the velocity of the peak of the [Fe\,{\sc
ii}] emission is closer to the velocity of the the optical emission
lines.  Pa$\beta$ is faint, but the normalized profile is
considerably broader than the [Fe\,{\sc ii}] emission profiles. The Gaussian FWHM
is 48.2~km\,s$^{-1}$ and the total profile width at 1\% of the peak is $\sim$75km\,s$^{-1}$.
The peak of the profile appears to be asymmetric.
In Fig. \ref{velocityfig2} we compare the Pa$\beta$ profile with two
other Paschen lines (Pa$\delta$ at 1004.9373 nm and Pa$_20$ at 839.2396
nm) from the UVES spectrum.  The absorption profile of Pa\,3-20 has
very broad wings and some emission. This emission becomes stronger for lower members
of the Paschen series.

\subsection{Optical [O\,{\sc i}] and [C\,{\sc i}] emission lines}

The only forbidden emission lines in the optical, UVES spectrum 
(Reyniers\,\cite{Reyniers02}) are the [O\,{\sc i}] lines at 5577.3387,
6300.304, 6363.777\,\AA\ and the [C\,{\sc i}] lines at 8727.12,
9824.13, and 9850.26\,\AA. We have plotted the strongest line of each
element in Fig. \ref{velocityfig1}. We notice
 that the optical [O\,{\sc i}] and [C\,{\sc i}] emission lines have a 
profile very similar to the H$_2$ emission line. 
Because the UVES spectra were obtained with the image slicer, we have
no spatial information. However based on their similar profiles, they seem to
originate in the shock which produces the H$_2$ emission.
In PNe, collisionally excited C$^0$ has also been observed to co-exist 
mainly with H$_2$, making it an invaluable probe of H$_2$ emitting regions (Liu et
al. \cite{Liu95}).  In old extended PNe, such
as the Ring nebula, the [O\,{\sc i}] 6300 \AA\ line is also found to
largely follow the near-infrared H$_2$ emission (Liu \& Barlow
\cite{Liu96}). 

The [C\,{\sc i}]\,$\lambda$\,9824, 9850\,\AA\ lines decay from the
same upper level, so their intensity ratio depends only on the
relative transition probabilities, and is purely determined by atomic
parameters.  Similarly, the [O\,{\sc i}]\,$\lambda$\,6300, 6363\,\AA\
line ratio is also purely determined by atomic parameters.  In the case of
collisional excitation, the observed [C\,{\sc i}] line ratio
$(I9824\,+\,I9850)\,/\,I8727$ and the [O\,{\sc i}] ratio
$(I6300\,+\,I6363)\,/\,I5577$ provide each a direct diagnostic of the
electron temperature, assuming the electron density is known.  Using
the extinction derived in Paper~I (R$_V=4.2$, E(B$-$V)\,$=\,1.78$), we
corrected the lines for extinction and calculated both ratios.  For
the [C\,{\sc i}] line ratio we obtain a value of 1.5\,$\pm$\,0.5.
For the [O\,{\sc i}] ratio we obtain a value of 6.0\,$\pm$\,1.0.  
The UVES slit width was 0\farcs7, orientated east-west, and, because of
the use of the slicer, no atmospheric dispersion correction could be
applied. The shift between the 5577 \AA\ and the 6363~\AA\ lines is
0\farcs2.  Because the object is extended, this probably has little
effect on the ratio. We checked whether the [C\,{\sc i}] lines are
affected by telluric absorption using the transition probabilities
calculated by Nussbaumer \& Rusca (\cite{Nussbaumer79}). The
$I9824\,/\,I9850$ theoretical line ratio is 0.338 and our 
value is within 3\% of this theoretical value. Hence 
in our spectra these lines are not affected by telluric absorption

In order to calculate the electron temperature from the [O\,{\sc i}] or
[C\,{\sc i}] line ratios, we need to assume an electron density.
Usually, the [N\,{\sc i}] doublet ratio $\lambda$~5200.42/5197.95~\AA\
is one of the very few density-sensitive diagnostics observable from
neutral species in the optical region. It is a good tracer of the
electron density in regions where [C\,{\sc i}] and [O\,{\sc i}] are
observed.  However the [N\,{\sc i}] lines are not detected.  This is
not due to ionization of nitrogen.  The ionization potential of N$^0$
(14.534 eV) is a bit larger than that of H$^0$ (13.62 eV) and O$^0$
(13.60 eV), and no [O\,{\sc ii}] or [N\,{\sc ii}] lines are observed
in the optical spectrum. Nitrogen doesn't seem to be underabundant
(Reyniers \cite{Reyniers02}) either.  Hence the absence of the
[N\,{\sc i}] lines indicates that the [N\,{\sc i}] lines are
collisionally de-excited in the high-density region where the
[C\,{\sc i}] and [O\,{\sc i}] lines originate.  This is due to the
relatively low critical densities of the [N\,{\sc i}] lines: about
4830 and 1160 cm$^{-3}$ at $T_{\rm e} = 10,000$ K for the upper levels
of the $\lambda$ 5198 and 5200 \AA\ lines, respectively (Zeippen
\cite{Zeippen82}, Berrington \& Burke \cite{Berrington81}).

For carbon, the critical densities of the 2p$^2$\,$^1$D$_2$ level from which the
$\lambda$\,9824\,\AA\ and 9850\,\AA\ transitions originate are
$4.8\,\times\,10^4$, $2.2\,\times\,10^4$, and $1.6\,\times\,10^4$
cm$^{-3}$ for $T_{\rm e}\,=\,1000,\,5000$, and 10,000\,K, respectively
(Liu et al. \cite{Liu95}) and $1.1\,\times\,10^7$\,cm$^{-3}$ for the
$^1{\rm S}_0$ level from which the $\lambda\,8727$\,\AA\ line originates
(Mendoza \cite{Mendoza83}, Berrington \& Burke \cite{Berrington81}).
Hence the low value derived from the [C\,{\sc i}] ratio 
indicates that the $^1$D$_2$ level population is quenched by electron
impacts, and that the electron density in the shocked region would be
higher than $10^5$\,cm$^{-3}$.  The critical density for the $^1{\rm D}_2$ level 
from which the $\lambda$\,6300\,\AA\ and 6363\,\AA\ lines
originate is $1.9\,\times\,10^6$\,cm$^{-3}$ at 10,000\,K and is
$1.1\,\times\,10^8$\,cm$^{-3}$ for the $^1{\rm S}_0$ level from which
the $\lambda\,5577$\,\AA\ line originates
 (Mendoza \cite{Mendoza83}, Berrington \& Burke \cite{Berrington81}).
The [O\,{\sc i}] line ratio is insensitive to electron densities below approximately $10^6$\,cm$^{-3}$ at
10,000\,K.

As we have no constraint on the electron density, we have used the
photo-ionization code Cloudy (version 07.10.15) last described by Ferland
(\cite{Ferland98}) to calculate a grid of [O\,{\sc i}] and [C\,{\sc i}] 
line ratios in function of $T_{\rm e}$ and $n_{\rm e}$.
The detection of strong collisionally excited
[C\,{\sc i}] and [O\,{\sc i}] line emission, but no [O\,{\sc ii}] line emission, 
requires that there must be a region which is
predominantly neutral, but hot enough for effective excitation of the
levels in [O\,{\sc i}], which implies $6000~\la~T_{\rm e}~({\rm
K})~\la~10,000$.  The electron density must be higher than $10^4$\,cm$^{-3}$ for the [N\,{\sc i}]
lines to be collisionally de-excited.
We considered $ 3.7\la{\rm log}(T_{\rm e}/{\rm K})\la4.0$ and $4.0\la{\rm log}(n_{\rm e}/{\rm cm}^{-3})\la8.0$,  
using steps of 0.1 dex and 0.25 dex respectively.  
The [O\,{\sc i}] ratio corresponds to electron densities of $3\,\times\,10^6$ to
$5\,\times\,10^7$\,cm$^{-3}$ for electron temperatures of 10,000\,K to 6,000\,K, respectively.  
At these electron temperatures the electron densities derived from the
[C\,{\sc i}] line ratio are $10^5 - 3\times10^{5}$\,cm$^{-3}$.  

In summary: There is no indication of ionized gas in the optical
spectrum of IRAS 16594-4656. The forbidden [O\,{\sc i}] and [C\,{\sc
i}] emission are collisionally excited and originate in a very high
density region ($3\,\times\,10^6~\la\,n_{\rm e}\,\la\,5\,\times\,10^7$\,cm$^{-3}$), in the same
shock which produces the H$_2$ emission.

\section{Discussion}
\label{discussion}

\subsection{Kinematics and morphology of the H$_2$ 1-0 S(1) emission.}

In order to investigate the 3-dimensional structure of the lobes, we
constructed a data cube using all existing data in H$_2$ 1-0 S(1).
First we rectified the H$_2$ 1-0 S(1) data at each position angle to
have a pixel scale of $0.085\arcsec$ pix$^{-1}$ in the spatial
dimension and of 1.263\,km\,s$^{-1}$ in the velocity dimension.  Then we
aligned the position-velocity frames using the location of the
continuum peak in the spatial dimension and the reference velocity in
the velocity dimension.  Next we spline-interpolated the surface
brightness as a function of position angle in each pixel in the
position velocity frame based on the original seven frames.  Finally
we projected the interpolated 3-dimensional data onto a 3-dimensional
cartesian grid.

\begin{figure*}
\mbox{\includegraphics*[width=18cm]{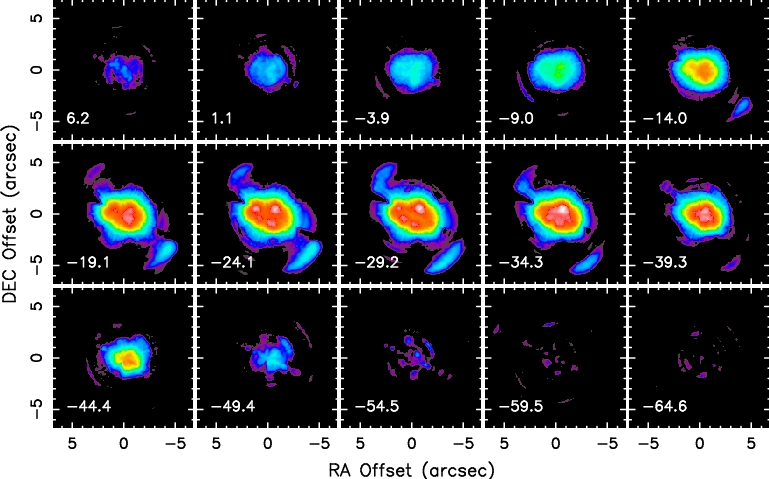}}
\\
\caption{\label{vcuts}The velocity channel maps of the reconstructed
 H$_2$ 1-0 S(1) data cube centered around the reference velocity of $-29.2$\,km\,s$^{-1}$. 
Each channel has a width of 5.05\,km\,s$^{-1}$ and its central velocity is indicated at the bottom 
left corner in\,km\,s$^{-1}$.}
\end{figure*}

Fig.\,\ref{vcuts} shows the velocity channel maps that are recovered
from the reconstructed data cube in H$_2$ 1-0 S(1). Each channel has a
width of 5.05\,km\,s$^{-1}$. Maps near the reference velocity indicate
that the H$_2$ emission is strong in both lobes, but stronger in the
western lobe than in the eastern lobe, which is also seen in the total
intensity map in Fig.\ \ref{slitpos}. They also indicate that there is
in fact less emission along the waist region of the shell.  This can also be seen 
from the position velocity maps at $7^{\circ}$ and $345^{\circ}$, but 
the velocity channel maps depict it more eloquently.  Thus, the overall
3-dimensional structure of the shell is such that the majority of the
H$_2$ emission is distributed as hollow bipolar lobes, one at the approaching
side and one at the receding side of the shell. 
The approaching side of the shell is elongated towards
the west while the receding side of the shell is elongated towards
the east.

\begin{figure}
\mbox{\includegraphics{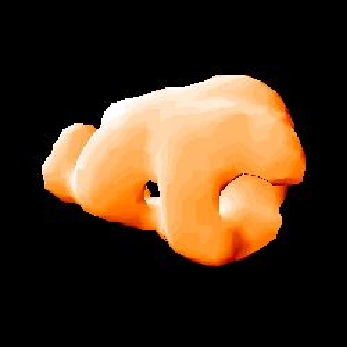}}
\caption{\label{modelfig}%
3-dimensional rendition of the shell structure of IRAS 16594$-$4656.
The surface represents an isointensity contour at roughly $30\%$ of the
 peak intensity.}
\end{figure}

Assuming a homologous expansion, i.e., an expansion with $v \propto r$, 
the velocity axis in the recovered data cube corresponds to a spatial
axis along the line of sight.  This allows us to visualize the
3-dimensional structure of the spatial distribution of H$_2$.  
Fig.\,\ref{modelfig} shows the 3-dimensional representation of the central
shell of IRAS 16594-4656, for which the surface contour has been
constructed based on the reconstructed data cube at $30\%$ of the peak
emission.  Fig.\ \ref{modelfig} shows the bottom right view of the
surface contour, illuminated obliquely from the southwest side by a
fiducial light source. The 3-dimensional volume shows the central
cavity and the hollowness of the shell along the
polar axis of the bipolar lobes, the western and eastern elongations
in the approaching and receding sides of the shell, and the holes on
the surface at the top and bottom of the equatorial regions.

\subsection{Shock properties}

In Paper\,I we concluded that the H$_2$ emission in IRAS 16594$-$4656
is mainly collisionally excited and that (1) the line ratios indicate
that the H$_2$ emission originates in a 20 to 30\,km\,s$^{-1}$ C-type shock in
$10^3$\,cm$^{-3}$ material (Le Bourlot et al. \cite{Lebourlot02}), (2)
we postulated that the C-shocks occur where the fast wind is funneled
through the EDE, which may harbor a magnetic field, but we noted that
the H$_2$ emission could also be present in the lobes, excited by
shocks caused by a molecular outflow impinging on the AGB envelope,
and (3) we noted that in principle H$_2$ and Fe$^+$ cannot coexist in
the same shocked region in substantial quantities.

In previous sections we showed that the H$_2$ emission is indeed
present at the edge of the EDE, but is more intense at the edge of the
lobes.  The H$_2$ emission appears to be excited in shocks between the
AGB ejecta and the post-AGB wind.

In this paper we found that the [O\,{\sc i}] and [C\,{\sc i}] emission
lines are very similar to the H$_2$ profiles in shape and
velocity. They may originate in the shock close to where the H$_2$
emission originates.  Their line ratios indicate a very high density
of $3\,\times\,10^6$ to $5\,\times\,10^7$ cm$^{-3}$ in the shock.  At this high
density the rotational temperature of 1440\,$\pm$ 80\,K derived in
Paper\,I corresponds to a 10 to 20 km\,s$^{-1}$ C-type shock into
$10^6$ to $10^7$ cm$^{-3}$ material (Le Bourlot et al. \cite{Lebourlot02}).  
However at such high densities, the
ionization of the metals may suffice to produce a J-type shock
(Smith \cite{Smith94}). If this were the case, the rotational temperature
determined in Paper\,I and the high density would also be consistent with a
J-type shock of 5 to 10 km\,s$^{-1}$ into $10^5$ to $10^6$\,cm$^{-3}$
material (Wilgenbus et al. \cite{Wilgenbus00}).  
At these low velocities and high densities the
J-type shock will not dissociate H$_2$, and the H$_2$ emission will
remain strong.

[Fe\,{\sc ii}] 1.64~$\mu$m is a cooling line of gas that is excited
predominantly by shocks.  It is a very good tracer of gas shocked by
energetic events in the absence of photoionization, which is the case for
IRAS~16594-4656.  [Fe\,{\sc ii}] emission cannot be excited in the presence of
C-shocks alone, as C-type shocks do not produce emission from ionized species
such as Fe$^+$.  [Fe\,{\sc ii}] emission is well modelled by fast dissociative
J-type shocks. In slow non-dissociative J-shocks, the  [Fe\,{\sc ii}] emission drops
rapidly with the shock velocity (Gredel \cite{Gredel94}). Hence
[Fe\,{\sc ii}] emission arises from fast dissociating and ionizing shocks, while
the H$_2$ emission arises from locations where the shock velocities are low
enough, not to cause destruction of H$_2$.
We see indeed that the H$_2$ and [Fe\,{\sc ii}] emission line profiles
and spatial extent are sufficiently different to confirm that they
mainly come from different regions.  The [Fe\,{\sc ii}] emission
appears to originate in a region close to the central star within the
hollow structure outlined by the H$_2$ emission.  The [Fe\,{\sc
ii}] emission is not located along the side walls of the hourglass, as
was considered as one of the possibilities in Paper\,I and is the case
in Hubble~12. Neither is it present as blobs in discrete shock fronts
in the lobes as is the case for M~1-92 (Davis et al. \cite{Davis05}).
The [Fe\,{\sc ii}] emission may be due to shocks in
the stellar wind or originate in a circumstellar or circumbinary disk.

\section{Conclusions}
\label{conclusions}

The H$_2$ emission is excited in slow 5 to 20 km\,s$^{-1}$
shocks into dense material at the edge of the lobes, caused by the
interaction of the AGB ejecta and the post-AGB wind.  The 3D
representation of the H$_2$ emission shows a hollow structure.  There
is less H$_2$ emission in the equatorial region.  The collisionally
excited [O\,{\sc i}] and [C\,{\sc i}] optical emission lines have a
similar profile compared to the extracted H$_2$ profile and appear to be
produced in the same shock.  They all indicate an expansion velocity
of $\sim$~8~km~s$^{-1}$ and the presence of a neutral, very high
density region of about $3\,\times\,10^6$ to $5\,\times\,10^7$~ cm$^{-3}$.  The
[Fe\,{\sc ii}] emission is not present in the lobes, but originates
close to the central star. It originates in fast shocks in the
post-AGB wind, or in a circumstellar or circumbinary disk. The
Pa$\beta$ emission also appears to originate close to the star. The
total width of the [Fe\,{\sc ii}] and Pa$\beta$ emission line profiles
is 62 and 75\,km~s$^{-1}$ respectively, at 1\% of the peak.

\begin{acknowledgements}

TU and PvH acknowledge support from the Belgian Science Policy Office in the framework of the IUAP5/36 project.  PvH acknowledges support from the Belgian Science Policy Office through grant MO/33/017. MR acknowledge financial support from the Fund for Scientific  Research - Flanders (Belgium). We thank. B. Hrivnak for stimulating discussions.

\end{acknowledgements}

\end{document}